\documentclass[12pt, a4paper]{article} %
\usepackage{graphicx} %
\usepackage[utf8]{inputenc} %
\usepackage{color} %
\usepackage{listings} %
\usepackage{hyperref}
\hypersetup{
    colorlinks,%
    citecolor=black,%
    filecolor=black,%
    linkcolor=black,%
    urlcolor=black
}
\usepackage{amsfonts} %
\usepackage{amsmath} %
\usepackage{amssymb} %
\usepackage{mathrsfs} %
\usepackage{latexsym} %
\usepackage{euscript} %
\usepackage{cite} %
\usepackage{mciteplus}
\DeclareTextSymbol{\degre}{OT1}{23}

\newcommand \degr{\!\degre\!\!} %

\newcommand \be{\begin{equation}} %
\newcommand \ee{\end{equation}} %
\newcommand \bea{\begin{eqnarray}} %
\newcommand \eea{\end{eqnarray}} %
\newcommand \ba{\begin{array}} %
\newcommand \ea{\end{array}} %

\title{\textsc{New polarimetric constraints\\on axion-like particles}}

\author{A.~Payez, J.R.~Cudell, and D.~Hutsem{\'e}kers\thanks{IFPA group, AGO Dept., U.~of Li\`ege, B4000 Li\`ege, Belgium.\newline e-mails: A.Payez@ulg.ac.be; JR.Cudell@ulg.ac.be; D.Hutsemekers@ulg.ac.be.}
}

\date{
}

\begin{document}

	\maketitle

	\begin{abstract}
		We show that the parameter space of axion-like particles can be severly constrained using high-precision measurements of quasar polarisations. Robust limits are derived from the measured bounds on optical circular polarisation and from the distribution of linear polarisations of quasars. As an outlook, this technique can be improved by the observation of objects located behind clusters of galaxies, using upcoming space-borne X-ray polarimeters.
	
	\end{abstract}

	\section{Introduction}

		From the theoretical point of view, an observation of ``axion-like particles'' (ALPs) would provide us with valuable information about theories beyond the standard model. Indeed, these spinless particles\hspace{1pt}---\hspace{1pt}with properties similar to those of the QCD axion, but with different masses and couplings\hspace{1pt}---\hspace{1pt}arise as testable predictions, as they are associated with the spontaneous symmetry breaking of new U(1) symmetries~\cite{Jaeckel:2010ni}.

		From the phenomenological point of view, such particles are also extremely appealing. First of all, depending on their masses and couplings, they can constitute a natural cold-dark-matter candidate; see, \textit{e.g.} Ref.~\cite{Sikivie:2009qn}. Moreover, as they can have a coupling with light similar to that of neutral pions, the existence of ALPs could help understand various astrophysical observations~\cite{Christensson:2002ig,*DeAngelis:2007dy,*Simet:2007sa,*Burrage:2009mj,*Fairbairn:2009zi,*Mirizzi:2009aj}.
		It is quite striking that these astrophysical hints seem to point towards the same kind of ALP, with a mass $m\lesssim 10^{-10}$~eV and a coupling to photons $g\sim10^{-11}$~GeV$^{-1}$, defining a whole new window of interest in the $(m,g)$ parameter space.

		To date, the most stringent constraints on ALPs are coming from astrophysics~\cite{Raffelt:2006cw}, and are mostly related to stellar dynamics: the existence of such particles would imply the opening of new channels to evacuate energy, which should have an impact on stellar evolution. A few years ago, the limit derived from the observation of the Horizontal Branch in the Hertzsprung--Russell diagram for globular clusters~\cite{Raffelt:1987yu,*Raffelt:1996bk} has been improved by CAST (the CERN Axion Solar Telescope), which tries to convert axions coming from the Sun~\cite{Zioutas:2004hi} back into (X-Ray) photons. Discussions of the energy loss in the SN1987A supernova have also lead to an even more severe upper bound on their couplings to photons~\cite{Brockway:1996yr}, which applies for nearly massless ALPs.

		The existence of such particles can also be probed using another consequence of their electromagnetic coupling.
		Until recently, it was hoped that a puzzling observation, involving quasars, could also be explained by low-mass ALPs. Indeed, in some huge ($\sim~\!\!\!1~$Gpc) regions in the sky, redshift-dependent correlations in the distribution of polarisation position angles for visible light from cosmologically distant quasars have been observed~\cite{Hutsemekers:1998,Hutsemekers:2001,Hutsemekers:2005,Jain:2003sg}. As one of the most interesting attributes of the mixing of axion-like particles with light is the induced modification of its linear and circular polarisations when it propagates in external magnetic fields~\cite{Sikivie:1983ip,Raffelt:1987im}, the presence of ALPs could have elegantly solved the problem~\cite{Jain:2002vx,Das:2004qka,Piotrovich:2008iy,Payez:2008pm,*Hutsemekers:2008iv,*Payez:2009kc,*Payez:2009vi,*Payez:2010xb,Agarwal:2009ic}. Given the present polarisation data~\cite{Hutsemekers:2010fw}, this scenario is now strongly disfavoured, already at the qualitative level~\cite{Payez:2011sh}: despite the use of a wave-packet treatment, the inclusion of fluctuations, and of more complex magnetic field morphologies, there is either no alignment, too much circular polarisation, or too much linear polarisation.
		Note that a recent claim of similar coherent orientations of polarisation in radiowaves~\cite{Tiwari:2012rr} would further disfavour an ALP mechanism~\cite{Payez:2012rc}. Henceforth, we shall not discuss the coherent alignment effect further.

		What we do in the present work is to use such polarisation data to constrain the properties of low-mass ALPs. The change of polarisation induced by the mixing is indeed a very specific prediction, and this is especially true when it comes to circular polarisation.
		Depending on mass and coupling of the ALPs, the mixing with light in external magnetic fields can be very efficient and contradict observations, as there is indeed little room for a modification of polarisation.
		Similar polarimetric constraints on these particles, based on linear polarisation, have already been discussed in the literature in various contexts, \textit{e.g.}~\cite{Jain:2002vx,Payez:2011mk,Gill:2011yp,Horns:2012pp}. Being as conservative as possible and using present quasar polarisation data in visible light, we show that one can derive new constraints on ALPs using supercluster magnetic fields, and narrow down the parameter-space region of astrophysical interest.

	\section{Polarisation of light and mixing with ALPs}\label{sec:mixing}

		The polarisation state of any light beam $\vec{\mathcal{E}}_r(z,t)=\mathcal{E}_{r_{1}}(z,t)\vec{e}_{1} + \mathcal{E}_{r_{2}}(z,t)\vec{e}_{2}$\hspace{1pt}---\hspace{1pt}where ($\vec{e}_{1},\vec{e}_{2}$) define the plane transverse to the propagation direction\hspace{1pt}---\hspace{1pt}is fully described by its Stokes parameters (which are averages over the exposure time). These can be written as
		\bea
                \left\{
                    \begin{array}{llll}
                        I(z) &=& \langle\mathcal{I}(z,t)\rangle \ \!= \langle \mathcal{E}^{\phantom{*}}_{r_{1}}\mathcal{E}^*_{r_{1}} + \mathcal{E}^{\phantom{*}}_{r_{2}}\mathcal{E}^*_{r_{2}} \rangle\\
                        Q(z) &=& \langle\mathcal{Q}(z,t)\rangle = \langle \mathcal{E}^{\phantom{*}}_{r_{2}}\mathcal{E}^*_{r_{2}} - \mathcal{E}^{\phantom{*}}_{r_{1}}\mathcal{E}^*_{r_{1}}\rangle\\
                        U(z) &=& \langle\mathcal{U}(z,t)\rangle = \langle \mathcal{E}^{\phantom{*}}_{r_{1}}\mathcal{E}^*_{r_{2}} + \mathcal{E}^*_{r_{1}}\mathcal{E}^{\phantom{*}}_{r_{2}}\rangle\\
                        V(z) &=& \langle\mathcal{V}(z,t)\rangle = \langle i(\mathcal{E}^{\phantom{*}}_{r_{1}}\mathcal{E}^*_{r_{2}} - \mathcal{E}^*_{r_{1}}\mathcal{E}^{\phantom{*}}_{r_{2}}) \rangle,
                    \end{array} \right.
                \eea
		and, by combining them, one introduces two intrinsic physical quantities: the degree of linear polarisation and the degree of circular polarisation, respectively
		\be
			p_{\mathrm{lin}}=\frac{\sqrt{Q^2+U^2}}{I}\qquad {\textrm{and}} \qquad	p_{\mathrm{circ}} = \frac{|V|}{I}.
		\ee

		Before we use polarisation data to derive new constraints on ALPs, let us recall~\cite{Payez:2011sh} that, whenever a mixing with axion-like particles takes place inside external magnetic fields, the polarisation of light changes. Even for initially unpolarised light, this mixing will lead to a modification of the linear polarisation (due to dichroism), while, for partially polarised light, some interplay between the circular and the linear polarisations can also occur (due to birefringence). These properties emerge from the fact that, given the direction of the transverse external magnetic field, ALPs only couple to a specific direction of polarisation of light, in a slightly different\footnote{Scalars only couple to photons with a polarisation perpendicular to the magnetic field, while pseudoscalars only couple to parallel ones.} way depending on the scalar or pseudoscalar character of the ALP. In this paper, we are going to particularise to the case of pseudoscalar ALPs but bear in mind that similar results hold for scalar ALPs.

		In regions of constant magnetic field, the evolution of the polarisation depends only on two dimensionless parameters~\cite{Payez:2011sh}. One of them is the mixing angle, which unambiguously gives the maximum amount of polarisation that can be achieved,
		\be
			\theta_{\mathrm{mix}} = \frac{1}{2} \textrm{atan}\left(\frac{2g\mathcal{B}\omega}{m^2 - {\omega_{\mathrm{p}}}^2}\right),\label{eq:thetamix}
		\ee
		and which involves the coupling constant between photons and pseudoscalars $g$, the transverse external magnetic field strength $\mathcal{B}$, the frequency of the light beam $\omega$, the ALP mass $m$, and the plasma frequency $\omega_{\mathrm{p}}$. The second one determines the wavelength of the oscillations of the Stokes parameters, and is the product
		\be
			\frac{\Delta\mu^2 L}{\omega} = \frac{\sqrt{{{(2g\mathcal{B}\omega)}^2 + (m^2 - {\omega_{\mathrm{p}}}^2)}^2}L}{\omega},
		\ee
		with $\Delta\mu^2$ the difference of the squared masses of the two new eigenstates of propagation, and $L$ the distance travelled. Note that this is very similar to what one obtains for neutrino oscillations.
\\
		
		Things are more complicated in more general magnetic fields, mostly because variations of the electron density and of the magnetic field (both in strength and direction) imply variations of the above dimensionless quantities. Nonetheless, the coupling constant and the magnetic field strength always appear together as a product in the equations.

		In order to tackle this problem, we shall discretise the continuous magnetic field and the electron density in a succession of zones, and work at the amplitude level for the electromagnetic and axion fields, ensuring continuity at each boundary.\footnote{Note that we neglect reflected waves, which have an amplitude of order $\left(\frac{\Delta\mu^2}{\omega^2}\right)$.} The Stokes parameters are computed at the end of the last zone.

		Finally, note that, due to the mixing with axions, even an initially unpolarised light beam will be slightly linearly polarised at the end of the first zone, so that one generally expects some amount of circular polarisation to be created in such a field, as a result of birefringence.

	\section{Polarisation data}

		\subsection{Properties of the full sample}

			An all-sky sample of 355 good-quality measurements of the linear polarisation of visible light from quasars\footnote{Here, ``quasar'' refers to ``high-luminosity AGN.''} has been defined in a series of papers~\cite{Hutsemekers:1998,Hutsemekers:2001,Hutsemekers:2005}, based on data obtained in Refs.~\cite{Stockman:1984qz,Berriman:1990qz,Moore:1984qz,*Brotherton:1998qz,*Hutsemekers:1998pp,*Schmidt:1999p,*Lamy:2000qz,*Sluse:2005,Smith:2002qz,Impey:1990qz,*Impey:1991qz,*Wills:1992qz,*Visvanathan:1998qz}.
			In that sample, only objects with a polarisation degree $p_{\mathrm{lin}} \ge 0.6\%$ and $p_{\mathrm{lin}} / \sigma_{p_{\mathrm{lin}}} \ge 2$ were considered. In the present work, we extend the sample to include low-polarisation objects with $p_{\mathrm{lin}} < 0.6\%$ measured with uncertainties $\sigma_{p_{\mathrm{lin}}} \le 0.3\%$. Since $p_{\mathrm{lin}}$ is positive definite, it is biased at low signal-to-noise ratio. A reasonably good estimator of the true (debiased) polarisation degree is computed using $\tilde{p}_{\mathrm{lin}} = (p^2_{{\mathrm{lin}}} - \sigma^2_{p_{\mathrm{lin}}})^{\frac{1}{2}}$ when $p_{\mathrm{lin}} > \sigma_{p_{\mathrm{lin}}}$, and $\tilde{p}_{\mathrm{lin}} = 0$ when $p_{\mathrm{lin}} \le \sigma_{p_{\mathrm{lin}}}$~\cite{Simmons:1985ub}. In the rest of this paper, we only consider $\tilde{p}_{\mathrm{lin}}$ when we discuss the linear polarisation.

			Circular polarisation measurements using a Bessell V-filter\footnote{This filter is centered around $\lambda=547.6$~nm, with a full width at half maximum of 113.2~nm.}~\cite{Bessell:1990hs} for some objects in the sample are available in~\cite{Hutsemekers:2010fw}, together with a compilation of previous measurements~\cite{Landstreet:1972so,Stockman:1984qz,Moore:1981circ,*Valtaoja:1993circ,*Takalo:1993circ,*Impey:1995circ,*Tommasi:2001a,Tommasi:2001b}, most of them in white light (with no filter\footnote{For instance, the photomultipliers used in~\cite{Landstreet:1972so} have a broad spectral response, from 185 to 930~nm. Note however that there is an atmospheric cutoff for wavelengths below 330~nm~\cite{Patat:2011atm}.}).  Out of 21 V-filter measurements reported in~\cite{Hutsemekers:2010fw}, all but two are compatible with no circular polarisation at $3\sigma$. The two positive detections concern highly linearly polarised blazars (both with $p_{\mathrm{lin}}>22\%$), which could be intrinsically circularly polarised~\cite{Hutsemekers:2010fw}, as in the case of a highly linearly polarised BL Lac object ($p_{\mathrm{lin}}>26\%$), for which a non-zero circular polarisation in white light has also been previously observed~\cite{Tommasi:2001b,Hutsemekers:2010fw}. These special cases set aside, light is essentially not circularly polarised, be it white or V-filtered.

		\subsection{Subsample and criteria used to obtain constraints}\label{sec:criteria}

			\begin{figure}
				\centering
				\includegraphics[height=6.825cm]{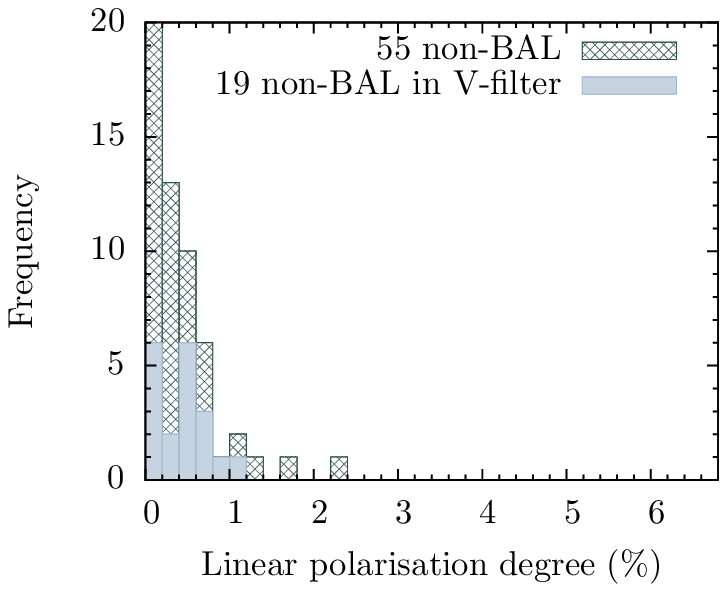}
				\includegraphics[height=6.825cm]{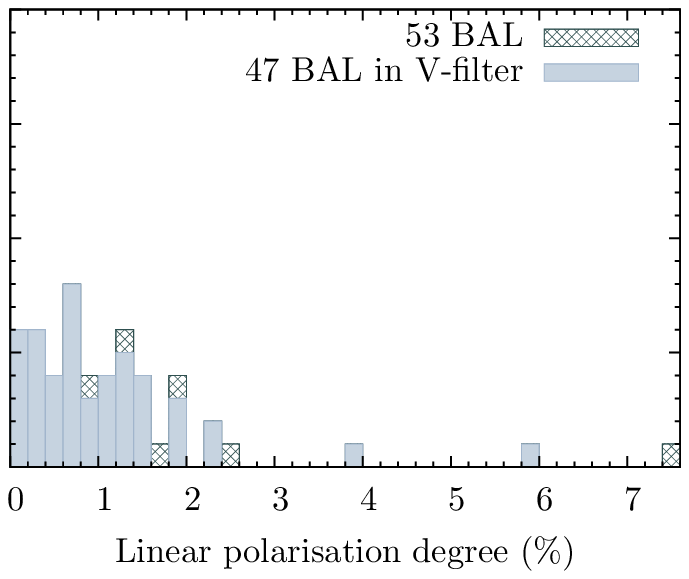}
				\caption{Comparison of the distributions of debiased linear polarisation between non-BAL quasars (\emph{left}) and BAL quasars (\emph{right}), for objects taken in the direction of region A1. Clearly, BAL quasars are often more linearly polarised than quasars without broad absorption lines.}
				\label{fig:balnonbal}
			\end{figure}

			Since our goal is to derive constraints from the polarisation induced by axion-like particles, we restrict our sample to spectroscopically defined classes of quasars known to have the smallest intrinsic polarisations.

			We discard the following objects from the sample of quasars with linear polarisation measurements: known radio-loud quasars (\textit{i.e.}, classified as such in the NASA/IPAC Extragalactic Database (NED) or objects from references~\cite{Impey:1990qz,*Impey:1991qz,*Wills:1992qz,*Visvanathan:1998qz}), near-infrared (2MASS) selected objects, in particular those from~\cite{Smith:2002qz}, and Broad-Absorption-Line (BAL) quasars (classified as such in NED). A comparison of non-BAL measurements to the polarisation distribution of BAL quasars is shown in Fig.~\ref{fig:balnonbal}, demonstrating that differences are significant; they are also preserved by any mechanism possibly affecting light on the line-of-sight, see Ref.~\cite{Hutsemekers:2001}.

			In the sample of circularly polarised quasars, we discard the two spectroscopically identified BL Lac objects, more prone to be intrinsically polarised.

			The reduced sample then mostly consists of radio-quiet / optically selected non-BAL quasars.  Such objects are known to be linearly polarised at most at the 1\% level~\cite{Stockman:1984qz,Berriman:1990qz}.

\begin{figure}
				\centering
				\includegraphics[height=6.825cm]{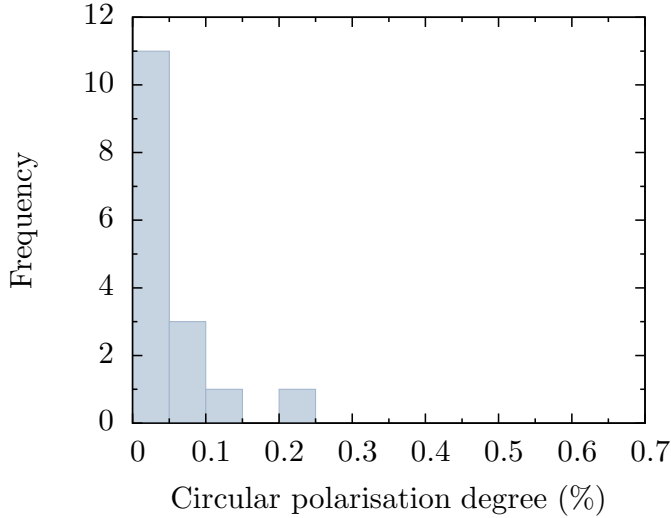}
				\caption{Distribution of the degree of circular polarisation for the central values of the 16 quasars in the direction of region A1, measured in the V-filter.}
				\label{fig:pcirc}
			\end{figure}

			As the observation of null circular polarisation is a much more stringent constraint for axion-like particles in V-filter than in white light~\cite{Payez:2011sh}, we focus on quasars located in the direction in which most of these V-filter data were taken, \textit{i.e.} in the direction of the so-called ``region A1''~\cite{Hutsemekers:1998,Hutsemekers:2001,Hutsemekers:2005}, \textit{i.e.} towards the center of the (local) Virgo supercluster. Keeping only objects with right ascensions between 168\degr{} and 218\degr, we are left with a final sample of 55 quasars with measured linear polarisation in white light or in V-filter and of 16 quasars with measured circular polarisation in V-filter (all compatible with zero at $3\sigma$). The distribution of the circular polarisation degree associated with the central values of the 16 circular polarisation measurements is shown in Fig.~\ref{fig:pcirc}.

			We know that these quasars intrinsically emit polarised light: otherwise we would simply not see a difference in the distribution of degrees of linear polarisation depending on the spectroscopic type. However, we cannot access the initial distribution of polarisation. For this reason, in order to avoid any overestimation of the final polarisation generated by the mixing, we work with initially unpolarised light.  We conservatively allow the observed linear polarisation of the quasars to be only due to the interaction with ALPs.
			Under this hypothesis, one can impose that the linear polarisation generated by the ALP-photon mixing model does not exceed the observed one.

			The idea is therefore to calculate the polarisation predicted by the mixing with ALPs inside external magnetic fields encountered on the way to Earth and to compare it with data.
			For both the linear and the circular polarisations, we thus compute the probability that the polarisation ($p^{\textrm{th}}$) due to ALPs is smaller than the observed one ($p^{\mathrm{obs}}$) as:
			\be
			P = N (p^{\mathrm{obs}} \geq p^{\textrm{th}}) / N_{\mathrm{total}},\label{eq:proba}
			\ee
			by taking the ratio of the number of measurements in bins with a polarisation higher than $p^{\textrm{th}}$, to the total number of measurements $N_{\mathrm{total}}$.

	\section{Model for supercluster magnetic fields}

		\subsection{Structure}\label{sec:struct}

			On the observational side, what is typically assumed in the literature for magnetic fields inside clusters or superclusters is a cell-like structure, see \textit{e.g.}~\cite{Vallee:2002,Vallee:2011,Xu:2005rb}, possibly with an additional weaker background field coherent over several cells. One usually considers a collection of cells of a given size, and lets the magnetic field direction change from cell to cell while keeping the same field strength $|\vec{{B}}_{\mathrm{cell}}|$,\footnote{Note that this kind of model is also what is usually used when one discusses magnetic fields beyond supercluster scales; see for instance~\cite{Mirizzi:2005ng,*Mirizzi:2006zy}. Such faint cosmological magnetic fields have also been used to obtain limits on ALPs~\cite{Horns:2012pp}.} which averages to a smaller value at the supercluster scale.
	
			In such a configuration, one sees different fields when looking in different directions. Because of this, and because the exact magnetic structure is not known inside the Virgo supercluster, we assume that the effect of these irregularities is equivalent to considering that light coming from different objects essentially passes through random fields.

			Additionally, we perform general tridimensional rotations of the magnetic field direction from cell to cell; it will thus pick up a longitudinal component most of the time. As mentioned in Sec.~\ref{sec:mixing}, only the projection of the magnetic field onto the tranverse plane is relevant for ALP-photon mixing. Hence, we allow values of the transverse magnetic field strength $\mathcal{B}$ much smaller than $|\vec{B}_{\mathrm{cell}}|$.

			We also consider fluctuations of the electron density from cell to cell and of the size of each cell.
			More precisely, we let these quantities fluctuate by 50\% up or down.
			Note that, as we allow for fluctuations of cell sizes, only the total distance $z_{\mathrm{tot}}$ is well-defined from one magnetic field realisation to another, not the number of cells, as we stop adding cells as soon as we reach $z_{\mathrm{tot}}$.

			Finally, note that the size of the supercluster ($\sim$ 20~Mpc) is small enough so that cosmological redshift can be neglected. As the mixing is more efficient at higher frequencies (see Eq.~\eqref{eq:thetamix}), redshift effects could only increase the polarisation produced in any case relevant to this study and make our bounds stronger.
			Moreover, as we only take into account the influence of the last magnetic field region, we again underestimate the amount of polarisation due to ALP-photon mixing.
			We thus stress that our constraints are conservative.

		\subsection{Field strength}\label{sec:Bstrength}

		Magnetic fields in galaxy clusters are well established~\cite{Clarke:2000bz,Giovannini:2003yn,Vallee:2011} and have typical strengths of several $\mu$G, although they can be as large as 40$~\mu$G. Our knowledge of these fields is sometimes refined enough that not only the field strength and the coherence lengths are known but also hints about the orientation are available~\cite{Pfrommer:2009hn}. In our case, even though region A1 is roughly centered on the Virgo cluster, most of the quasars are seen through a larger domain, and we need to model the magnetic field at the scale of the supercluster.

		There is evidence for a sizeable magnetic field in the surroundings of galaxy clusters~\cite{Kronberg:2007wa}, and some results are also available for superclusters.
		For instance, it has been inferred from rotation measures in radio wavelengths~\cite{Vallee:2002,Vallee:2011} that the structure of the magnetic field within the local supercluster plane (centered on the Virgo cluster, the direction we focus on) can be described as a collection of $\sim$2~$\mu$G magnetic field zones coherent over $\sim$100~kpc. If one interprets these data with a magnetic field coherent over the supercluster scale, or assumes much larger coherence lengths, one then obtains field strengths about 5 times smaller.
		A similar larger-scale magnetic field is also considered in Ref.~\cite{Xu:2005rb}, which reports upper values for typical magnetic field strengths in the Hercules and in the Perseus-Pisces superclusters of the order of $0.3\pm0.1~\mu$G (over 800~kpc) or $0.4\pm0.2~\mu$G (over 400~kpc).

	\section{Minimal constraints}

	\subsection{Method}\label{sec:method}

		Our constraints on ALPs are derived iteratively for different points in the $(m,g)$ parameter space: a given couple of parameters corresponds to a given hypothetical axion-like particle. For fixed ALP parameters, we generate random configurations, defined as the set of transverse field strengths and directions, of cell sizes, and of electron densities.
		For a given value of the frequency $\omega$, we have the solutions for the two electric fields of the radiation after a propagation in each ``patch'' of such a configuration~\cite{Payez:2011sh}. 
		We use initially unpolarised light beams (as discussed in Sec.~\ref{sec:criteria}), calculate the four Stokes parameters at the end of the last magnetic field cell, and compare the final polarisation with data.

		Now, as explained in~\cite{Payez:2011sh}, the expected circular polarisation due to ALP-photon mixing can be reduced when one considers wave packets, or averages over frequencies. This effect, though typically small in V-filter, is taken into account when deriving our limits.

		To implement this, for a single light beam, we completely generate a random configuration. We then integrate over $\omega$ the monochromatic results for the Stokes parameters at the end of the last zone of magnetic field, with the spectral response of the V-filter as a weighting function. This corresponds to one iteration: it gives the values of the Stokes parameters through the V-filter for a given source at the end of one configuration.

		After that, we compare the generated linear and circular polarisations with the data histograms\footnote{The results are not affected by the presence of white light data for linear polarisation.}, associate to each of them a probability given by Eq.~\eqref{eq:proba}, respectively $P_{\mathrm{lin}}$ and $P_{\mathrm{circ}}$, and obtain the final probability $\mathcal{P}$ to be compatible with data for this particular configuration by multiplying these two individual probabilities. We then repeat the procedure with a new random configuration.

		We do this many times (5000 times) to minimise the influence of statistical fluctuations, and average the value of $\mathcal{P}$ over the configurations as we do not have access to the actual configuration for each measurement. We thus integrate it out and give the average probability for a given ALP $(m,g)$ not to exceed the observed polarisation in a random configuration. Then we start over for a new couple $(m,g)$.

		Finally, we summarise this information by saying that parameters are ``excluded at 1$\sigma$'', when the average probability for this particle to produce too much polarisation compared with data is 68.3\%; and similarly for 2$\sigma$ (95.5\%) and 3$\sigma$ (99.7\%).
		In Appendix~\ref{app}, we show two theoretical distributions of polarisation, associated with points excluded at 2$\sigma$ and 3$\sigma$; let us stress that this is only for illustration, as the technique described above to obtain constraints does not directly involve such distributions.

	\subsection{Results}

		\begin{figure}[h]
			\includegraphics[width=\textwidth]{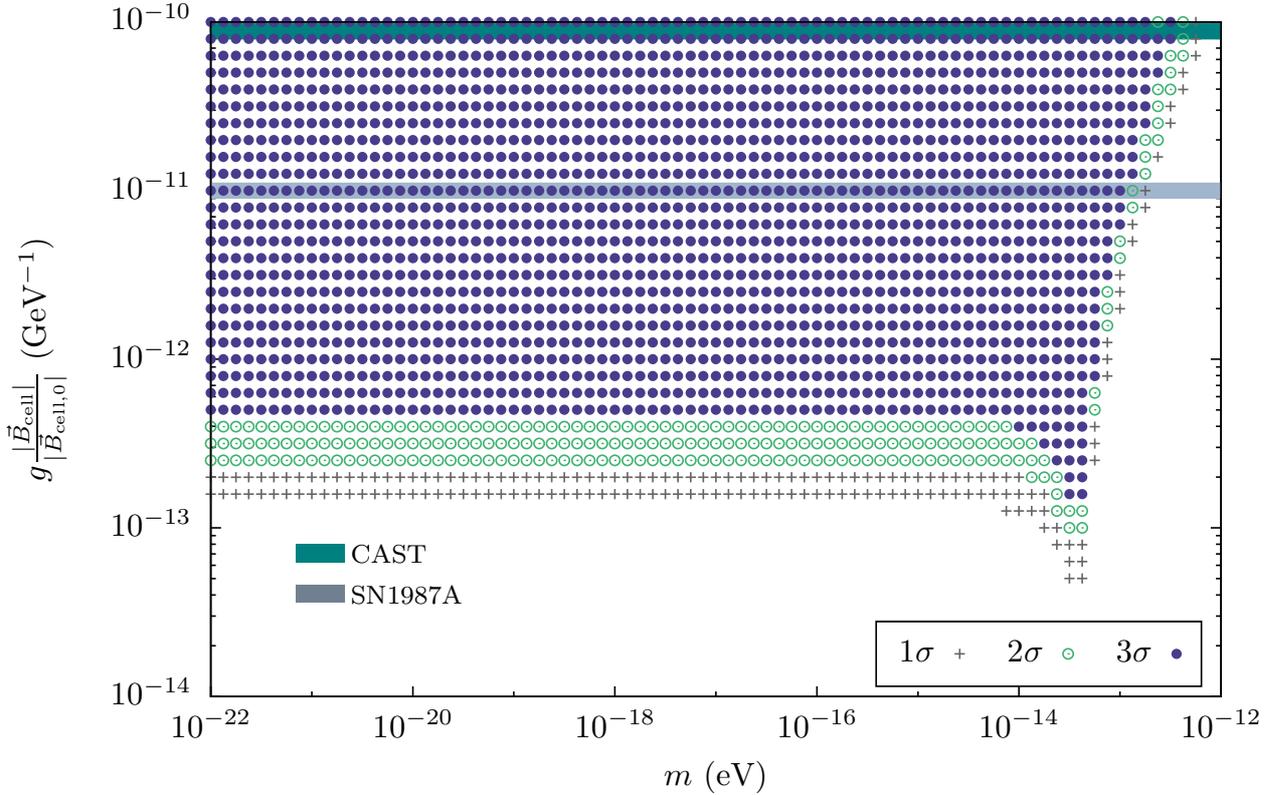}
			\caption{Exclusion plot. The total distance considered here is $z_{\mathrm{tot}}=10$~Mpc; the average cell size is 100~kpc; the norm of the total magnetic field in each cell is $|\vec{B}_{\mathrm{cell,0}}|=2~\mu$G; the average electron density is $n_{\mathrm{e}}=10^{-6}$~cm$^{-3}$ (corresponding to a plasma frequency $\omega_{\mathrm{p}}=3.7\times10^{-14}$~eV).
			\label{fig:exclusion}
			}
		\end{figure}

		The results we obtain are shown in Fig.~\ref{fig:exclusion}. The upper limits obtained by CAST and from the energy-loss considerations associated with SN1987A are also shown, and confirmed for nearly massless ALPs.

		Let us first stress a very important point: as there is only one physical magnetic field scale in the problem, namely the total field strength inside a cell $|\vec{B}_{\mathrm{cell}}|$, this exclusion plot remains the same for any other value of the magnetic field inside the cells. Indeed, as stressed in Sec.~\ref{sec:mixing}, only the product of the magnetic field with the coupling constant $g$ enters the equations. Therefore, while there are uncertainties about the magnetic field strength, as discussed in Sec.~\ref{sec:Bstrength}, the exclusion limit on $g$ can be rescaled once our knowledge of the magnetic field inside the supercluster improves.

		In Fig.~\ref{fig:exclusion}, we observe two features: a dip and a plateau for small ALP masses. Both can be understood from the expression of the mixing angle, Eq.~\eqref{eq:thetamix}. The position of the dip is determined by $\omega_{\mathrm{p}}$; for ALP masses very close to the plasma frequency, the mixing is maximal, and so is its effect on polarisation. As for the plateau, for ALP masses much smaller than the plasma frequency, the mixing angle is essentially independent of $m$:
		\be
			\theta_{\mathrm{mix}}\stackrel{(m\lll\omega_{\mathrm{p}})}{=}\frac{1}{2} \textrm{atan}\left(\frac{2g\mathcal{B}\omega}{- {\omega_{\mathrm{p}}}^2}\right);
		\ee
		this exclusion thus holds even for massless ALPs.
		Now, the fact that higher values of the coupling $g$ are more constrained is natural, as the mixing is more efficient and, thus, more polarisation is produced.
		
		Here, for the average electron density inside the supercluster, we have used $n_{\mathrm{e}} = 10^{-6}$~cm$^{-3}$ (which gives a plasma frequency\footnote{The plasma frequency is given by $\omega_{\mathrm{p}}= 3.7\times10^{-14}\sqrt{{n_{\mathrm{e}}}/{10^{-6}~\textrm{cm}^{-3}}}~\textrm{eV}$.} $\omega_{\mathrm{p}}=3.7\times10^{-14}$~eV), as in previous works related to ALP-photon mixing in similar conditions, \textit{e.g.}~\cite{Das:2004qka,Burrage:2008ii}.
		Nonetheless, the properties of the intrasupercluster medium are not well-known, including the value of the average electron density.
		Searches for gas in the Shapley supercluster~\cite{Molnar:1998sc} have bounded $n_{\mathrm{e}}$ to be less than $5\times10^{-6}$~cm$^{-3}$, and a subsequent tentative detection, for the local supercluster~\cite{Boughn:1999sc}, gave $n_{\mathrm{e}}\approx2.5\times10^{-6}$~cm$^{-3}$. More recent constrained N-body simulations of the local supercluster in the direction we are looking at (at galactic latitudes $b>30\degr{}$ and distances of about 10~Mpc) lead to values for the plasma frequency between $\omega_{\mathrm{p}}=1.7\times10^{-14}$~eV and $\omega_{\mathrm{p}}=5.5\times10^{-14}$~eV for matter overdensities between $\delta_{\mathrm{g}}=0$ and $\delta_{\mathrm{g}}=10$~\cite{Kravtsov:2002ac}.

		Now, if we take an extreme case and allow the intrasupercluster electron density to be one order of magnitude larger on average than what we have considered in Fig.~\ref{fig:exclusion} (\textit{i.e.} allow fluctuations, see Sec.~\ref{sec:struct}, up to values as large as $n_{\mathrm{e}}=15\times10^{-6}$~cm$^{-3}$), we have checked that, while our limits would then change, they would still improve the current bounds on ALPs. For smaller values of the electron density $n_{\mathrm{e}}<10^{-6}$~cm$^{-3}$, our limits are stable, and would in fact be slightly more stringent. 

		We have checked the stability of our constraints under the change of the total magnetic field size $z_{\mathrm{tot}}$. For total magnetic field sizes of 5~Mpc and 20~Mpc, the exclusion plot obtained essentially does not change its shape compared to Fig.~\ref{fig:exclusion}, but is shifted along the $y$-axis. For instance, the $2\sigma$-limit we obtain for nearly massless ALPs is $g\lesssim2.5\times10^{-13}$~GeV$^{-1}$ for $z_{\mathrm{tot}}=10$~Mpc, and becomes $g\lesssim3.2\times10^{-13}$~GeV$^{-1}$ and $g\lesssim2\times10^{-13}$~GeV$^{-1}$ for $z_{\mathrm{tot}}=5$~Mpc and $z_{\mathrm{tot}}=20$~Mpc respectively.
\bigskip

		We have also checked that our constraints are stable: a) under the addition of a uniform background field (typically weaker: $|\vec{B}_{\mathrm{uniform}}|\sim0.4~\mu$G), which would allow some correlation between cells over the supercluster scale;\footnote{Then, of course, one cannot rescale the limits derived for $|\vec{B}_{\mathrm{cell}}|$ and $|\vec{B}_{\mathrm{uniform}}|$ changed independently.} b) if we use the histograms of linear polarisation for non-BAL quasars measured in white light and V-filter, or exclusively in V-filter to define our probability; c) if we include the measurements of circular polarisation for the two BL Lac objects; d) if we do not allow for fluctuations of the size of the cells and of the electron density from cell to cell.

	\section{Conclusions and outlooks}

		This paper extends previous proposals to use polarisation data to constrain ALP parameters~\cite{Jain:2002vx,Payez:2011mk} by using a realistic magnetic field and an average over frequencies. We have presented constraints derived simultaneously from linear and circular polarisation measurements, sticking to a conservative approach where we allowed the observed polarisation of non-BAL quasars to be only due to ALPs.
		Using reported properties of the intrasupercluster medium, we have showed that current bounds on nearly massless ALPs are improved.
		Indeed, for a large portion of the parameter space region of astrophysical interest, the polarisation produced by the mixing with such particles in this medium would be too large.

		While our constraints can be straightforwardly rescaled when our knowledge of the magnetic field strength improves, the weak point of the method is of course the uncertainties on the electron density and on the field strength, and more reliable constraints could be obtained using galaxy clusters where these properties are better determined. As the plasma frequency is much higher at the cluster level, an effect in visible light would however be suppressed, except at the resonance. Using polarisation data in X-rays instead, one could take advantage of galaxy clusters to constrain ALP properties; this could be achieved with instruments such as the polarimeter onboard of the GEMS satellite~\cite{Swank:2010gems,*Feroci:2011}, which will operate between 2 and 10~keV. Along with other authors~\cite{Bassan:2010ya,*Mena:2011xj}, we therefore stress that the polarisation properties of very-high-energy photons is a promising tool to search for ALPs.

	\section*{Acknowledgements}

	A.~P. thanks Davide Mancusi for many useful discussions on technical issues, and \'Eric Gosset for interesting exchanges regarding statistics. D.~H. is Senior Research Associate at F.R.S-FNRS; A.~P. is an IISN researcher. This research has made use of the NASA/IPAC Extragalactic Database (NED) which is operated by the Jet Propulsion Laboratory, California Institute of Technology, under contract with the National Aeronautics and Space Administration.

\appendix

\section{Illustration of theoretical distributions}\label{app}

	We show distributions for the linear and the circular polarisations obtained after 5000 different magnetic field realisations. We reproduce what is obtained for two points taken from Fig.~\ref{fig:exclusion}, both for an ALP mass $m=10^{-20}$~eV, but for different values of the coupling.

\begin{figure}
			\centering
			\includegraphics[height=6.cm]{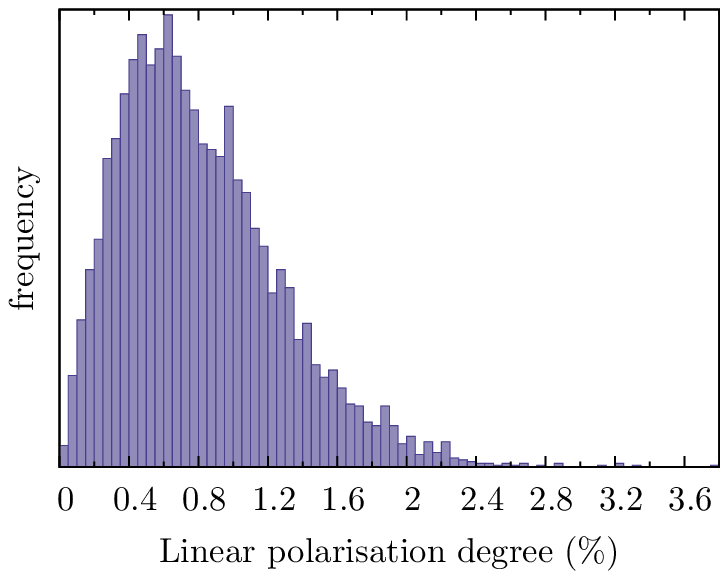}
			\includegraphics[height=6.cm]{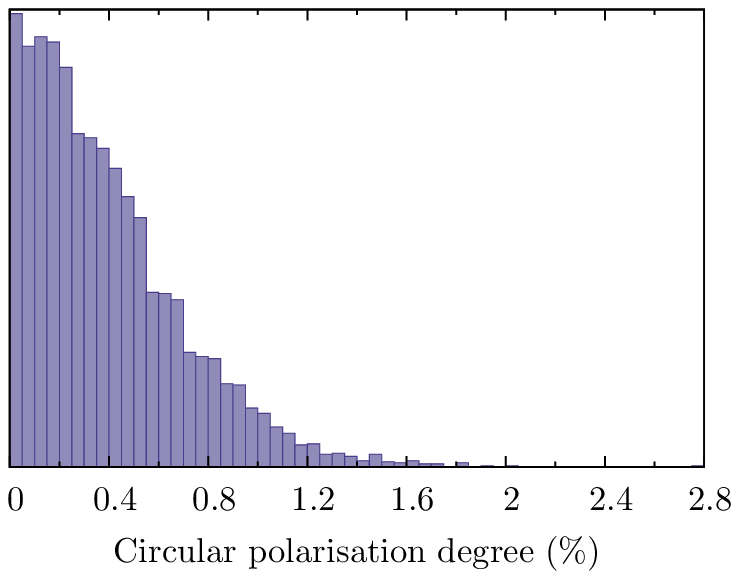}
			\caption{Theoretical distributions of polarisation for 5000 randomly generated configurations for initially unpolarised light. Here, $m=10^{-20}$~eV and $g=2.5\times10^{-13}$~GeV$^{-1}$; parameters and fluctuations are the same as in Fig.~\ref{fig:exclusion}.}
			\label{fig:2s}
		\end{figure}

\begin{figure}
			\centering
			\includegraphics[height=6.cm]{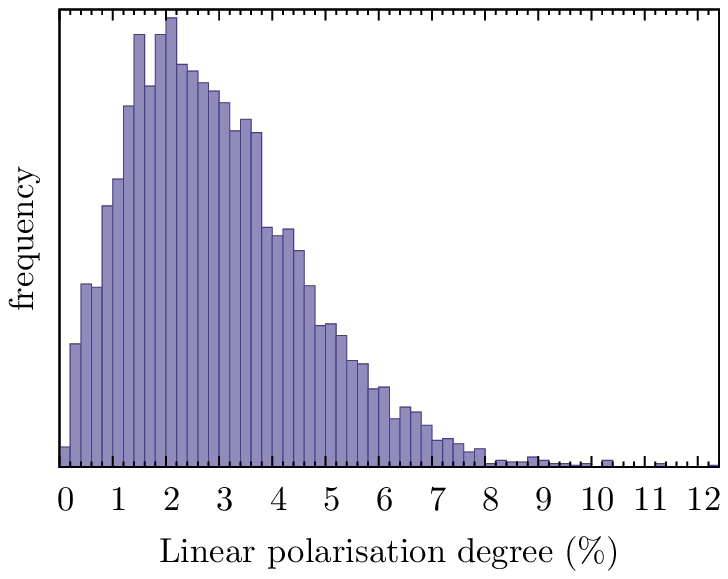}
			\includegraphics[height=6.cm]{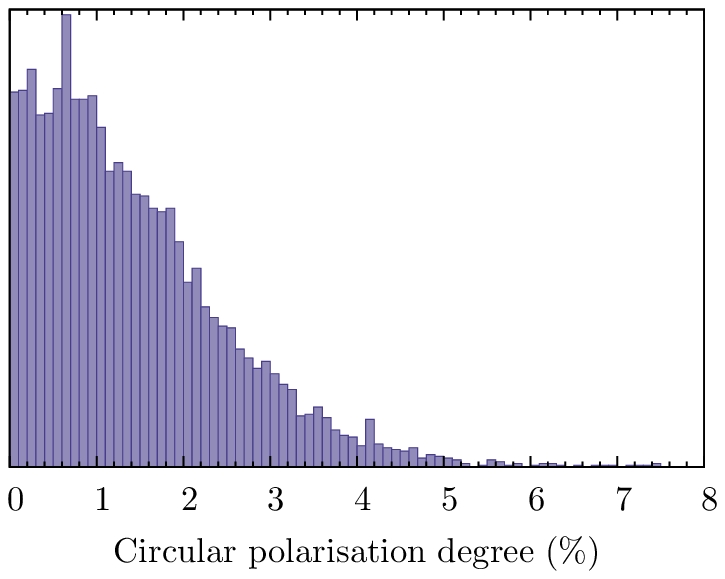}
			\caption{Theoretical distributions of polarisation for 5000 randomly generated configurations for initially unpolarised light. Here, $m=10^{-20}$~eV and $g=5\times10^{-13}$~GeV$^{-1}$; parameters and fluctuations are the same as in Fig.~\ref{fig:exclusion}.}
			\label{fig:3s}
		\end{figure}

	In Fig.~\ref{fig:2s}, we show the theoretical distribution of polarisations obtained for initially unpolarised light for a coupling $g=2.5\times10^{-13}$~GeV$^{-1}$, which is the lower bound of parameters excluded at 2$\sigma$ for nearly massless ALPs.
	In Fig.~\ref{fig:3s}, we show the same for a coupling $g=5\times10^{-13}$~GeV$^{-1}$, \textit{i.e.} the 3$\sigma$ lower bound for nearly massless ALPs.

	In both case, we see that ALP-photon mixing produces much more polarisation than observed (Figs.~\ref{fig:balnonbal} and~\ref{fig:pcirc}); in particular, the circular polarisation is very large. For values of the coupling close to the current bound from SN1987A, the linear and circular polarisations go above 50\% of polarisation in some configurations while we would expect at most around 1\% for the linear polarisation and a very tiny circular polarisation.

\newpage
\small

\bibliographystyle{modMminimalist}
\bibliography{alexbib}

\end{document}